# Block Based Medical Image Watermarking Technique for Tamper Detection and Recovery

Eswaraiah Rayachoti[1] and Sreenivasa Reddy Edara[2]

[1] CSE, VVIT
Guntur, Andhra Pradesh, India

[2] CSE, Acharya Nagarjuna University
Guntur, Andhra Pradesh, India

**Abstract**
In this paper, we propose a novel fragile block based medical image watermarking technique for embedding data of patient into medical image, verifying the integrity of ROI (Region of Interest), detecting the tampered blocks inside ROI and recovering original ROI with less size authentication and recovery data and with simple mathematical calculations. In the proposed method, the medical image is divided into three regions called ROI, RONI (Region of Non Interest) and border pixels. Later, authentication data of ROI and Electronic Patient Record (EPR) are compressed using Run Length Encoding (RLE) technique and then embedded into ROI. Recovery information of ROI is embedded inside RONI and information of ROI is embedded inside border pixels. Results of experiments conducted on several medical images reveal that proposed method produces high quality watermarked medical images, identifies tampered areas inside ROI of watermarked medical images and recovers the original ROI.

## 1. Introduction

Exchange of medical images between hospitals located at remote places has become a natural practice of modern times. This exchange of medical images inflicts two restraints for the medical images: the information has not been changed by unauthorized users and there should be evidence that the information belongs to the correct patient [1]. On the other hand transmission of patient data and his medical image separately through commercial networks like internet results in excessive transmission time and cost. Watermarking is one of the techniques used to deal with the above two concerns.

Watermarking techniques have been classified into two categories namely spatial domain and frequency domain. This classification is based on the medium used for concealing the data in an image. In spatial domain watermarking techniques [7, 10, 11, 13], data is embedded directly into host image while data is inserted into transformed host image in frequency domain watermarking techniques [8, 9, 14]. Another categorization of watermarking technique is reversible and irreversible. In reversible watermarking technique [9, 14, 17, 18], the host image can be recovered exactly at receiver side from watermarked image. Accurate recovery of host image is not possible in case of irreversible watermarking techniques [8, 15]. Reversible watermarking is more suitable for medical images [2].

Four types of watermarking methods are developed to protect digital images: Robust watermarking [3], Fragile Watermarking [4], Semi-Fragile Watermarking [5] and Hybrid Watermarking [6]. Robust watermarking methods are used for copyright protection of digital images as it is difficult to remove robust watermarks from digital images. Robust watermarks sustain intentional or unintentional attacks like scaling, compression, cropping and so on. Fragile watermarking techniques are best for checking authentication of digital images. Any modification or tampering removes fragile watermark from watermarked image. So, absence of watermark indicates that image has been tampered. Semi-Fragile watermarks survive only unintentional attacks. Hybrid watermarks are the amalgamation of fragile and robust watermarks. Here, robust watermarks are used for privacy control and fragile watermarks are used for the integrity control of the digital image.

Most of the medical images contain two parts called ROI and RONI. From diagnosis point of view ROI part is more important. Care should be taken while hiding data into ROI part so that visual quality will not be degraded. At the same time any tampering to ROI has to be identified and the original ROI has to be recovered in order to avoid misdiagnosis and retransmission of medical image. The recovery data of ROI is generally embedded into RONI [10, 11, 13, 14, 15, 16, 19]. When any tamper is detected inside ROI of received watermarked medical image then the tampered area of ROI is replaced with the recovery data embedded inside RONI.

In this paper, we are proposing a novel block based fragile medical image watermarking technique to achieve the following objectives.

1. Identifying the presence of tampers inside ROI.
2. Recovering the original ROI when it is tampered.
3. Detecting tampers inside ROI and recovering original ROI using minimal size authentication and recovery data.
4. Avoiding the process of checking ROI of the watermarked medical image for the presence of tampers when the ROI is not tampered.
5. Embedding EPR of patient into the medical image.

The rest of the paper is organized as follows. Section 2 covers review literature, proposed method is explained in section 3; results are illustrated in section 4 and finally conclusion in section 5.

## 2. Literature Review

Number of watermarking techniques has been developed for detecting tampers in the ROI or in the entire watermarked medical image and recovering the original ROI or the entire medical image. Zain *et al*. [7] proposed a block based scheme, the medical image is segmented into 8×8 blocks and then a mapping is established between the blocks for embedding the recovery information of each block into its corresponding mapped block. Later, each block is further divided into four sub blocks of 4×4 size each and then a 9-bit watermark is generated for each sub block. The generated 9 bit watermark of each sub block is embedded into LSBs of first 9 pixels of the sub block in the corresponding mapped block. At receiver's end, the watermarked medical image is divided into blocks of 8×8 size and then the mapping between the blocks is calculated as done in embedding procedure. Later, each block is further divided into four sub blocks of 4×4 size and then a 2-level detection scheme is applied for detecting tampered blocks. This 2-level detection scheme identifies tampered blocks. Where Level-1 detection is applied on sub-blocks of blocks and level-2 detection is applied on blocks. When a tampered block is detected, the corresponding mapped block is identified and then recovery data embedded in mapped block is extracted. This recovery data is used to replace the pixels in tampered block. Major drawbacks of this method are: 1) if both block A and its mapped block B are tampered then it is not possible to recover original image, 2) this method is not using any authentication data for the entire medical image to check directly whether the image is tampered. So, all blocks in the image have to be checked one after other to detect the presence of tampers. This checking process leads to wastage of time when the image is not tampered, 3) there is no provision for embedding EPR of patient into the medical image.

Wu *et al*. [8] developed two block based methods. In the second method, JPEG bit-string of the selected ROI is generated and then is divided into fixed length segments. Later, the medical image is divided into blocks and then hash bits are calculated for each block excluding the block with ROI. This hash bits are used as authentication data of the blocks. In each block of image, hash bits of the block and one segment of JPEG bit-string of ROI are both embedded using robust additive watermarking technique. Then all blocks are combined to get watermarked medical image. At receiver's end, the watermarked medical image is divided into blocks as done in embedding procedure. From each block, hash bits of the block and a segment of JPEG bit-string are both extracted. For each block, hash bits are calculated and then compared with the extracted hash bits to check whether the block is tampered or not. If the block with ROI is identified as tampered then the JPEG bit-string segments extracted from all blocks are used to recover the ROI. Disadvantages of this method are: 1) it is not possible to get original ROI as JPEG bit-string of ROI is used to recover ROI when it is tampered with, 2) this method requires more number of calculations to generate recovery data of ROI and embedding it into all blocks of medical image, 3) the size of authentication data is large; for each block 150 bits are used, 4) there is no provision for embedding EPR of patient into the medical image.

Chiang *et al.* [9] proposed two block based methods based on symmetric key cryptosystem and modified difference expansion (DE) technique. The first method has the ability to recover the whole medical image, where as the second method has the ability to recover only ROI of medical image. In the first method, the medical image is divided into 4×4 size blocks and then average of each block is calculated. Later, the averages of all blocks are concatenated and then encrypted using two symmetric keys k1 and k2 in order to increase the degree of security. Then, Haar wavelet transform is applied on all blocks to identify smooth blocks. The encrypted averages of all the blocks are embedded in the identified smooth blocks. At the receiver's end, the embedded data is extracted from watermarked image and then decrypted using the keys k1 and k2 to get the averages of all blocks. Later, averages are calculated for all blocks and then compared with extracted averages to detect tampered blocks. When a tampered block is detected then the pixels in tampered block are replaced with the extracted average of that block. The second method is same as the first method except that the bits of pixels in blocks of ROI are embedded instead of averages of all blocks in entire image. Pitfalls of these schemes are: 1) in the second method the size of authentication and recovery data is large; 128 bits for each block in ROI, 2) the two methods require more time for embedding data into medical image as all blocks of the medical image have to be transformed into frequency domain and then smooth blocks have to be identified for embedding data, 3) the two methods are not using any authentication data for the entire ROI or the entire image to check directly whether the ROI or the entire image is tampered. So, all blocks in the ROI or in the entire image have to be checked one after another to detect the presence of tampers. This checking process leads to wastage of time when the image is not tampered, 4) there is no provision for embedding EPR of patient into the medical image.

Liew *et al.* [10, 11] developed two reversible block based methods. In the first method, the medical image is segmented into two regions: ROI and RONI. Later, ROI and RONI are divided into non overlapping blocks of size 8×8 and 6×6 respectively. Then, a mapping is formed between blocks of ROI to embed recovery information of each block into its mapped block. Each block in ROI is mapped to a block in RONI. This mapping is used to embed LSBs of pixels in a ROI block into its mapped RONI block. Then, the method implemented by Zain *et al.* [7] is applied only on ROI part of the medical image for detecting tampers inside ROI and recovering original ROI. The LSBs of pixels inside ROI are replaced with its original bits that were stored inside RONI to make the scheme reversible. Second method is same as first method except that the removed LSBs of pixels in blocks of ROI are compressed using Run Length Encoding technique before embedding into RONI blocks. Drawbacks of the two methods are: 1) if both block A and its mapped block B inside ROI are tampered then it is not possible to recover original ROI, 2) the two methods are not using any authentication data for the entire ROI to check directly whether the ROI is tampered. So, all blocks in the ROI have to be checked one after another to detect the presence of tampers. This checking process leads to wastage of time when the ROI is not tampered, 3) there is no provision for embedding EPR of patient into the medical image.

Memon *et al.* [12] implemented a hybrid watermarking method. In this method, the medical image is segmented into ROI and RONI. Then, a fragile watermark is embedded into LSBs of ROI. RONI is divided into blocks of size N×N and then a location map indicating embeddable blocks is generated. A robust watermark is embedded into embeddable blocks of RONI using Integer Wavelet Transform (IWT). Later, the location map is embedded into $LL_3$ of each block using LSB substitution method. Finally, ROI and RONI are combined to get watermarked image. At receiver's end, the watermarked medical image is segmented into ROI and RONI. Then, the robust watermark is extracted from RONI and is used for checking authentication of image. Fragile watermark is extracted from ROI and checked visually to know presence of tampers inside ROI. Two disadvantages of this method are: 1) there is no specification of how the original ROI is recovered when the ROI is tampered, 2) the time complexity of this method is more as it has to generate location map before embedding data.

Agung *et al.* [13] developed a reversible method for medical images whose ROI size is more compared to size of RONI. In this method, the original LSBs of all pixels in medical image are collected and then LSB in each pixel is set to zero. Later, the medical image is segmented into ROI and RONI regions. Then, ROI and RONI are divided into blocks of size 6×6 and 6×1 respectively. A mapping is formed between blocks of ROI for storing recovery information of each ROI block into its mapped ROI block. The removed original LSBs are compressed using RLE technique and then embedded into 2 LSBs of 6×1 blocks in RONI. At receiver's end, the watermarked medical image is segmented into ROI and RONI as done in embedding procedure. Then, the method proposed by Zain [7] is applied only on ROI part to detect tampers inside ROI and recover original ROI. The original LSBs that were embedded in RONI are extracted and then restored to their positions to get the original medical image. This method has the same drawbacks as with methods proposed by Liew *et al*. [10, 11].

Qershi *et al.* [14] developed a reversible ROI based watermarking scheme. At sender's end, the medical image is segmented into ROI and RONI. Later, data of patient and hash value of ROI are both embedded into ROI using technique developed by Gou et al. Compressed form of ROI, average values of blocks inside ROI, embedding map for ROI, embedding map for RONI and LSBs of pixels in a secrete area of RONI are embedded into RONI using the technique of Tian. Finally, information of ROI is embedded into LSBs of pixels in secrete area. At receiver's end, ROI information is extracted from secrete area and is used to identify ROI and RONI regions. From the identified RONI region compressed form of ROI, average values of blocks inside ROI, embedding map of ROI, embedding map of RONI and LSB of pixels in secrete area are extracted. Using the extracted location map of ROI, patient's data and hash value of ROI are extracted from ROI. Then, hash value of ROI is calculated and compared with extracted hash value. If there is a mismatch between the two hash values then the ROI is divided into 16×16 blocks. For each block, the average value is calculated and compared with the corresponding average value in the extracted average values. If they are not equal then the block is marked as tampered and replaced by the corresponding block of the compressed form of ROI. Two disadvantages of this method are: 1) extracting the embedded data from RONI without knowing the embedding map of RONI, 2) use of compressed form of ROI as recovery data for the ROI.

Qershi *et al.* [15] proposed a scheme based on two dimensional difference expansion (2D-DE). At sender's end, the medical image is divided into three regions: ROI pixels, RONI pixels and border pixels. Later, the concatenation of patient's data, hash value of ROI, bits of pixels inside ROI and LSBs of border pixels is compressed using Huffman coding and then embedded into RONI using 2D-DE technique. This embedding generates a location map which will be concatenated with information of ROI and then embedded into LSBs of border pixels. At receiver's end, from border pixels in the watermarked medical image both information of ROI and location map are extracted. Using this ROI information, ROI and RONI are identified. The extracted location map is used to extract patient's data, hash value of ROI, bits of pixels inside ROI and LSBs of border pixels from RONI. The process for detecting tampered blocks is same as the one used in [14]. Each tampered block is replaced by the corresponding block of pixels in the extracted ROI. The LSBs of border pixels are replaced using the extracted LSBs from RONI. A major drawback of this scheme is it is applicable to only the medical images whose ROI size is very less (up to 12% of size of entire image).

Qershi *et al.* [16] developed a hybrid ROI-based method. At sender's end, the medical image is divided into three regions: ROI, RONI and border pixels. Later, patient's data and hash value of ROI are embedded inside ROI using modified DE technique. The ROI location map along with compressed form of ROI and average intensities of blocks inside ROI are then embedded into RONI using DWT technique. Then, size of watermark that is inserted into RONI and ROI information are embedded inside border pixels using the same DWT technique. At receiver's end, ROI information is extracted from border pixels and is used to identify ROI and RONI regions. Compressed form of ROI, average intensities of blocks in ROI and location map of ROI are extracted from the identified RONI region. Using the extracted location map of ROI, patient's data and hash value of ROI are extracted from ROI. The procedure for detecting tampered blocks and recovering ROI is same as in [14]. Two disadvantages of this method are: 1) use of compressed form of ROI as recovery information for the ROI, 2) applicable to only images whose size is at least 512×512.

Deng *et al.* [17] developed a region-based tampering detection and recovering method based on reversible watermarking and quad-tree decomposition. In this method, original image is divided into blocks with high homogeneity using quad-tree decomposition and then a recovery feature is calculated for each block using linear interpolation of pixels. The recovery features of all blocks are embedded as first watermark using invertible integer transformation. Quad-tree information as second layer watermark is embedded using LSB replacement. In the authentication phase, the embedded watermark is extracted and the original image is recovered. The similar linear interpolation technique is utilized to get each block's feature. The tampering detection and localization can be achieved through comparing the extracted feature with the recomputed one. The extracted feature can be used to recover those tampered regions with high similarity to their original state. One drawback of this scheme is exact original image cannot be recovered when it is tampered.

## 3. Proposed Method

To achieve the above mentioned objectives, we propose a medical image watermarking technique in this paper.

### 3.1 Division of Medical Image

In a medical image, the ROI is the most important part for making diagnosis. A medical image may contain several disjoint ROI areas and may be in different shapes. The ROI parts are marked by a physician or by a clinician interactively. Each ROI area is represented by an enclosing polygon. The enclosing polygon is characterized by the number of vertices and their coordinates. In proposed method, the medical image is segmented into three regions of pixels: ROI pixels, RONI pixels and border pixels as shown in Fig. 1. In present work, we use medical images containing a single ROI. The proposed method can also be used with medical images containing multiple ROI areas. In this method, the outer three lines of pixels in the image are indicated as border.

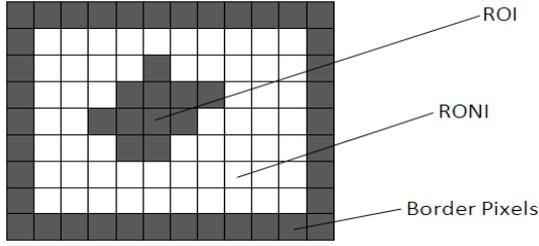

Fig. 1 Division of medical image into three regions.

### 3.2 Hash value of ROI

After selecting the ROI, the hash value of the ROI is calculated using the cryptographic hash function MD5. This function generates a unique code for any input and is a one way function. Determining the input from the code generated by MD5 is not possible. The calculated hash value of ROI is used to authenticate ROI.

### 3.3 Run Length Encoding

After calculating hash value of ROI, the ROI is divided into non overlapping blocks of size 4×4 and then average value is calculated for each block. A watermark is generated by concatenating hash value of ROI, LSBs of pixels inside ROI and EPR of patient. This generated watermark is compressed using RLE. RLE is a simple lossless compression technique and is used to reduce the size of watermark. Original data can be reconstructed exactly from the compressed data. In this technique, if a bit is repeating for number of times in sequence then that bit sequence is replaced by a count value and the bit. As an example, the binary data 000001111110000000 will be interpreted as five 0's, six 1's, seven 0's and it is coded as (101, 0), (110, 1) and (111, 0). The original binary data containing 18 bits is compressed to 12 bits. The compressed watermark is encrypted using a secret key k1 to provide security. The resultant watermark is embedded into LSBs of ROI pixels.

### 3.4 Mapping between blocks of ROI and RONI

After embedding watermark in LSBs of ROI pixels, RONI is divided into non overlapping blocks of size 3×3 pixels. Assuming that the number of blocks inside ROI is less than the number of blocks inside RONI, for each block in ROI the corresponding mapped block in RONI is identified using Eq. (1).

$$B_{RONI} = \lfloor (k \times B_{ROI}) \bmod N_b \rfloor + 1 \qquad (1)$$

where $N_b$ is the number of blocks in ROI, $B_{RONI}$ is block number in RONI, k is a secrete key and is a prime number between 1 and $N_b$, $B_{ROI}$ is block number in ROI. After mapping each ROI block to a RONI block, the average value of each ROI block is embedded inside the corresponding mapped RONI block.

Now, the detailed embedding algorithm is explained as follows.

### 3.5 Embedding Algorithm

1. Segment the original medical image into three regions: ROI pixels, RONI pixels and border pixels.
2. Calculate hash value (h1) of ROI using MD5.
3. Divide the pixels inside ROI into non overlapping blocks of size 4×4 each.
4. For each ROI block, calculate average value and use it as authentication and recovery data of that block.
5. Collect Least Significant Bits of all pixels inside ROI and denote this collection as B.
6. Represent the characters in EPR of patient using ASCII code and then get binary equivalent of it, E.
7. Generate watermark w by concatenating h1, B and E.
8. Compress watermark w using RLE compression technique to generate $w_{comp}$.
9. Encrypt the watermark $w_{comp}$ using a secret key k1.
10. Embed the bits of encrypted watermark into LSBs of pixels inside ROI.
11. Divide RONI into non overlapping blocks of size 3×3 each.
12. Assuming that the number of blocks in ROI is less than the number of blocks in RONI, map each block in ROI to a block in RONI using Eq. (1).
13. For each ROI block, calculate average intensity value and then embed into LSBs of first 8 pixels in mapped RONI block.
14. Encrypt the bits indicating the information of ROI using secret key k1.
15. Embed the encrypted bits into the LSBs of border pixels.

### 3.6 Extraction Algorithm

1. Extract the encrypted bits from the LSBs of border pixels in watermarked medical image.
2. Decrypt the extracted bits to get information of ROI.
3. Identify ROI pixels and RONI pixels in watermarked medical image.

4. Extract the encrypted watermark from the LSBs of pixels inside ROI.
5. Decrypt the extracted watermark to get $w_{comp}$.
6. Decompress the $w_{comp}$ to obtain the hash value (h1) of ROI, LSBs (B) of pixels inside ROI and EPR (E) of patient.
7. Replace the LSBs of pixels inside ROI with the bits in B.
8. Calculate hash value (h2) of the ROI using MD5.
9. Compare h1 with h2. If h1=h2 then the ROI is authentic and the extraction process ends.
10. If h1≠h2 then the ROI is not authentic and is tampered. Proceed to next step to detect tampered blocks inside ROI and recover original ROI.
11. Divide ROI and RONI into blocks of size 4×4 and 3×3 respectively. For each ROI block identify the mapped RONI block using Eq. (1) as in embedding procedure. For each ROI block, calculate average intensity and then compare it with the average intensity extracted from LSBs of first 8 pixels in the corresponding mapped RONI block. If they are not equal then mark the block as tampered and replace the pixels in this block with the extracted average value.

## 4. Experimental Results

We developed a MATLAB program for testing the performance of the proposed method. For conducting experiments, we used around hundred 8-bit grayscale medical images of different sizes and modalities like CT scan, MRI scan and Ultrasound. Out of these hundred images, 35 medical images are of CT scan, 40 medical images are of MRI scan and 25 medical images are of Ultrasound. Peak Signal to Noise Ratio (PSNR) and Weighted Peak Signal to Noise Ratio (WPSNR) [20] are used to measure the distortion in the generated watermarked medical images.

Higher value of PSNR and WPSNR designates less distortion in the watermarked image. Mean Structural SIMilarity index (MSSIM) [21] metric is used to measure the similarity between the original and the watermarked medical image. MSSIM value is between -1 and 1. Value 1 of MSSIM indicates that the original and watermarked images are similar. Visual degradation in the watermarked image is measured using the Total Perceptual Error (TPE) [22] metric. Lower value of TPE indicates less degradation in the watermarked image.

Some of the medical images used in our experiments are shown in Fig. 2. All images are resized to 256×256 and patient data of 0.5 KB size is embedded inside ROI. A rectangular shaped ROI is considered in each medical image for simulating the proposed method. Fig. 2 also shows the watermarked images generated after embedding watermark into original images and the watermark extracted or reconstructed medical images. There is no significant visual difference between the original, watermarked and watermark extracted medical images. Table 1 illustrates the results obtained after embedding watermark into the three medical images shown in Fig. 2. Table 2 depicts the average results obtained by watermarking the hundred medical images used in our experiments. Results shown in Tables 1 and 2 indicate that the proposed method works well for different modalities of medical images.

A medical image watermarking technique is effective if the PSNR value of watermarked and reconstructed medical image is greater than 40 dB [23]. In the proposed method, The PSNR and WPSNR values of watermarked and reconstructed medical images are above 40 dB. The perceived change in the structural information of the watermarked medical images is insignificant as the MSSIM values for all modalities of images are near to 1. Similarly, the low average TPE values show less visual degradation in the watermarked medical images.

The intruders are prevented from getting information of ROI by encrypting it before embedding inside border pixels. If an attacker identifies the ROI region and gets the LSBs of pixels inside ROI then he cannot do anything with that data as is encrypted by a secret key. Some of the state-of-the-art techniques [9, 10, 11, 13] are not using any authentication data, like hash value of ROI to check directly whether the ROI is tampered or not. So, all blocks inside ROI have to be checked one after the other to detect the presence of tampers. This checking process leads to wastage of time when the watermarked medical image is not tampered. Such wastage of time is not incurred in the proposed method as it is using hash value of ROI to directly check whether the ROI is tampered.

To test the performance of proposed method in terms of detecting tampered blocks inside ROI and recovering original ROI, we induced a tamper inside ROI of the watermarked medical images as shown in Fig. 3. Proposed method identified the tamper inside ROI and recovered original ROI.

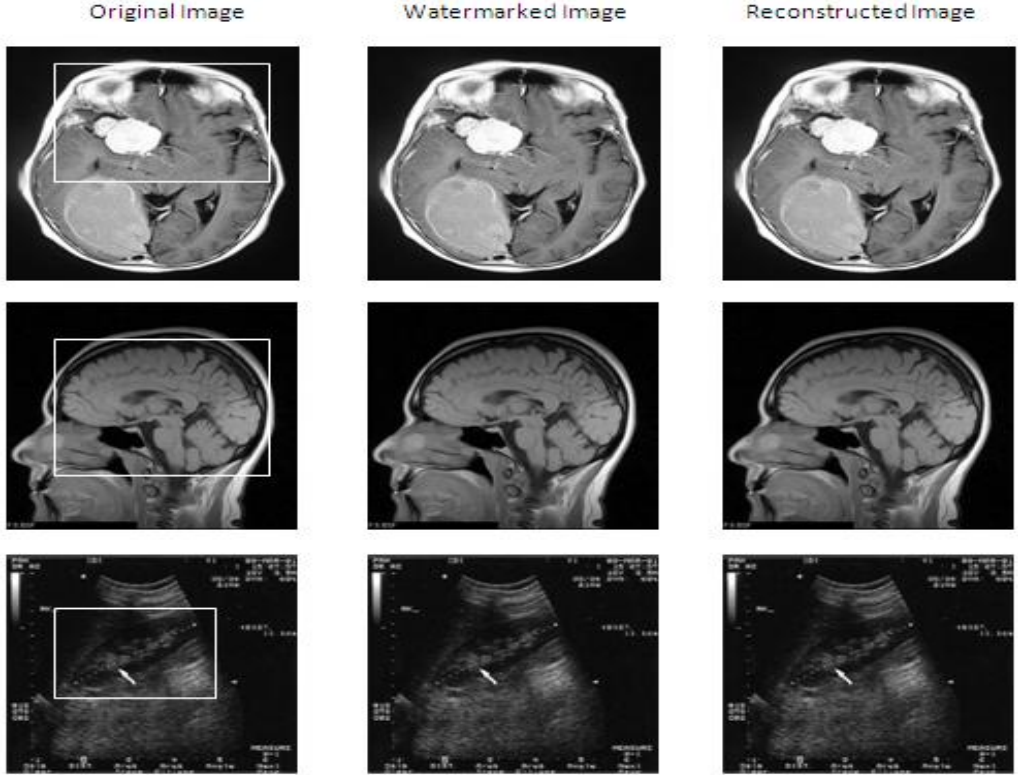

Fig. 2. Original, watermarked and reconstructed medical images. From top to bottom: CT scan, MRI scan and Ultrasound images.

Table 1: Results of embedding watermark into medical images of different modalities

| Modality | Size of ROI | Size of w (bits) | Size of $w_{comp}$ (bits) | Number of blocks in ROI | PSNR | WPSNR | MSSIM | TPE |
|---|---|---|---|---|---|---|---|---|
| CT | 200×192 | 42752 | 36348 | 2400 | 52.27 | 54.35 | 0.9347 | 0.0612 |
| MRI | 132×176 | 27584 | 23440 | 1452 | 57.34 | 58.13 | 0.9635 | 0.0445 |
| US | 104×128 | 17664 | 13072 | 832 | 60.56 | 63.21 | 0.9923 | 0.0221 |

Table 2: Performance of the proposed method

| Modality of Image | Average PSNR | Average WPSNR | Average MSSIM | Average TPE |
|---|---|---|---|---|
| CT Scan | 51.38 | 53.14 | 0.9216 | 0.0604 |
| MRI Scan | 54.26 | 56.89 | 0.9714 | 0.0468 |
| Ultrasound | 58.76 | 60.79 | 0.9851 | 0.0201 |

The reconstructed medical images are shown in Fig. 4. In a medical image, the LSB of pixels inside RONI and border are generally zero. So, the LSB of pixels inside RONI and border are set to 0 after extracting embedded data from them.

For testing the capability of proposed method in detecting tampers at multiple locations inside ROI and recovering original ROI, we modified pixels at number of locations inside ROI of watermarked medical images as shown in Fig. 5. Proposed method detected all the tampers inside ROI and recovered original ROI. Fig. 6 shows the reconstructed medical images. Some of the reviewed schemes [10, 11, 13] cannot recover ROI when tampers are induced by attackers at multiple locations inside ROI. Table 3 depicts the comparison between proposed method and the previously developed block based methods for tamper detection and recovery.

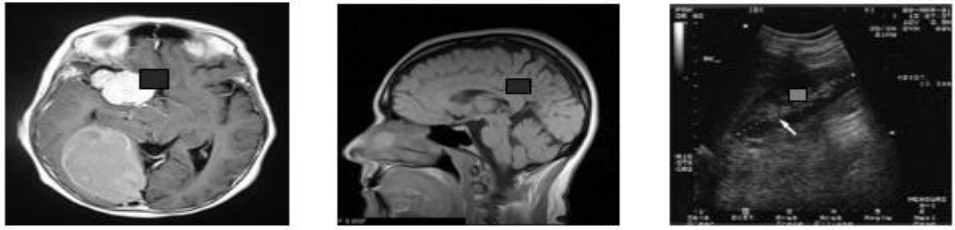
Fig. 3. Watermarked medical images (from left to right: CT scan, MRI scan and Ultrasound) with a tamper inside ROI.

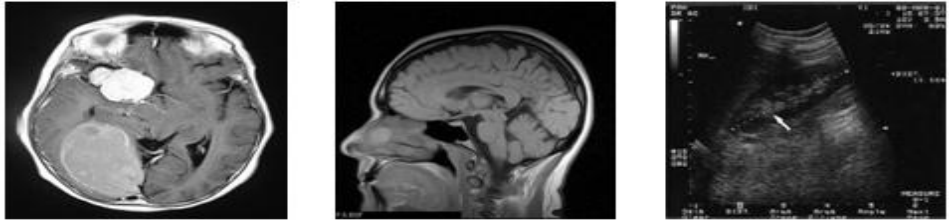
Fig. 4. Recovered medical images (from left to right: CT scan, MRI scan and Ultrasound).

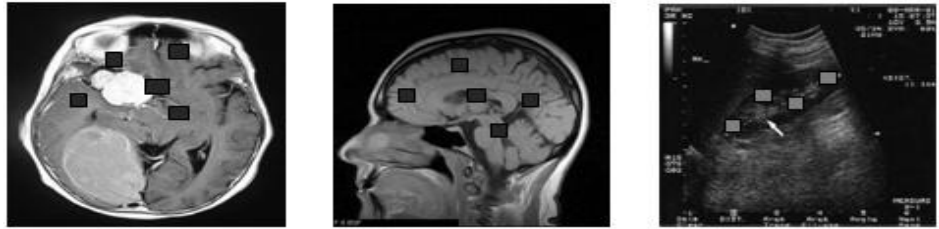
Fig. 5. Watermarked medical images (from left to right: CT scan, MRI scan and Ultrasound) with tampers at different locations inside ROI.

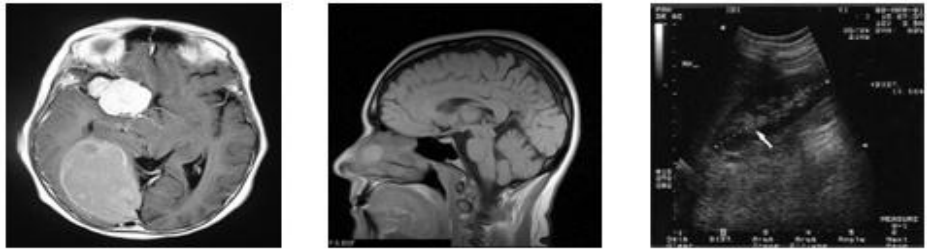
Fig. 6. Recovered medical images (from left to right: CT scan, MRI scan and Ultrasound).

Proposed method is developed based on the assumption that the intruders generally try to modify only the significant part, ROI, in the medical images during their transmission. So, identifying changes inside ROI and recovering original ROI must be done before using the medical image for diagnosis and to avoid misdiagnosis. One of the limitations of proposed method is: the RONI and border parts are not recovered exactly as LSBs of all pixels inside RONI and border are set to 0 after extracting embedded data from them. This limitation does not affect the efficiency of the proposed method as RONI and border parts of medical images are not significant for making diagnosis decisions. Another limitation is: no security for the ROI recovery data that is embedded inside RONI. If the ROI is tampered then the original ROI can be recovered only when the RONI and border part of the watermarked medical image are not attacked by any noise or not modified by intruders or not processed by common image manipulation operations.

## 4. Conclusions

Proposed medical image watermarking method produces high quality watermarked medical images. The watermarked medical images look more similar to original medical images as PSNR, WPSNR values of watermarked medical images are above 50dB and MSSIM values are above 0.93. Proposed method can be used with medical images whose ROI part is up to 62% of entire image. Proposed method uses only 8 bit authentication and recovery data for each 4×4 block inside ROI. It identifies and localizes tampers inside ROI and recovers original ROI. When the extracted hash value of ROI matches with recalculated hash value of ROI, then the proposed method do not check the blocks inside ROI for detecting the presence of tampers. Computational complexity of proposed method is less as it uses simple mathematical calculations for generating authentication and recovery data, detecting tampered blocks inside ROI and recovering original ROI.

For future enhancement, we try to extend the method for medical images whose pixels are represented using 10 or 12 or 16 bits and also to sustain common attacks, reduce embedding distortion inside ROI and recover the pixels inside ROI with their original bits instead of with average of pixels.

Table 3: Comparison between reviewed schemes and proposed scheme

| *Scheme* | *ROI-based* | *Size of authentication and recovery data* | *Is there any provision for embedding EPR* | *Recovery of ROI/image when it is tampered* |
|---|---|---|---|---|
| Zain | No | 9 bits for each 4×4 block | No | Not possible if a block and its mapping block both are tampered |
| Wu | Yes | 150 bits for each block | No | Possible, but with only compressed form of ROI |
| Chiang | Yes | 128 bits for each 4×4 block | No | Yes |
| Liew[10,11] | Yes | 9 bits for each 4×4 block | No | Not possible if a block and its mapping block both are tampered |
| Memon | Yes | - | Yes | No |
| Agung | Yes | 9 bits for each 3×3 block | No | Not possible if a block and its mapping block both are tampered |
| Qershi[15] | Yes | 128 bits for each 4×4 block | Yes | Yes |
| Qershi[16] | Yes | - | Yes | Possible, but with only compressed form of ROI |
| Deng | No | 8 bits for each block | No | Yes |
| Our method | Yes | 8 bits for each 4×4 block | Yes | Yes |